\documentclass[aps,amsmath,amssymb,amsfonts,twocolumn]{revtex4}
\usepackage{epsfig}
\usepackage{graphicx}
\usepackage{subfigure}
\usepackage{dcolumn}
\usepackage{bm}
\usepackage{amsthm}
\usepackage{enumitem}
\usepackage{slashed}
\usepackage{braket}
\usepackage{amsmath}
\usepackage{mathtools}
\usepackage{bbold}

\makeatletter
\def\l@subsubsection#1#2{}
\makeatother

\newcommand{\calE}{\mathcal{E}}

\newcommand{\calH}{\mathcal{H}}

\newtheorem*{axiom}{Macrostate axiom}
\newtheorem*{axiom_refinement}{Refinement to the macrostate axiom}

\begin{document}

\title{A microscopic derivation of the quantum measurement postulates}

\author{Vyacheslav Lysov}
\email{vyacheslav.lysov@oist.jp}
\affiliation{Okinawa Institute of Science and Technology, 1919-1 Tancha, Onna-son, Okinawa 904-0495, Japan}

\author{Yasha Neiman}
\email{yashula@icloud.com}
\affiliation{Okinawa Institute of Science and Technology, 1919-1 Tancha, Onna-son, Okinawa 904-0495, Japan}

\date{\today}

\begin{abstract}
In the mid-19th century, both the laws of mechanics and thermodynamics were known, and both appeared fundamental. This was changed by Boltzmann and Gibbs, who showed that thermodynamics can be \emph{derived}, by applying mechanics to very large systems, and making simple statistical assumptions about their behavior. Similarly, when Quantum Mechanics (QM) was first discovered, it appeared to require two sets of postulates: one about the deterministic evolution of wavefunctions, and another about the probabilistic measurement process. Here again, the latter is derivable from the former: by applying unitary evolution to large systems (apparatuses, observers and environment), and making simple assumptions about their behavior, one can derive all the features of quantum measurement. We set out to demonstrate this claim, using a simple and explicit model of a quantum experiment, which we hope will be clear and compelling to the average physicist.
\end{abstract}

\maketitle
\newpage

\section{Levels of understanding} \label{sec:intro}

In this paper, we claim that the so-called ``quantum foundations'' literature has in fact made progress on the foundations of QM. This is contrary to the message that the field itself is broadcasting, and to its perception among general physicists. To understand our claim of progress and its contentiousness, it is useful to compare the situation with that in thermodynamics. 

In thermodynamics, we can discern three levels of understanding. Level 1 is that of Carnot: knowing the concepts of temperature, heat and entropy, and the equations that govern them, but lacking any underlying microscopic picture. Level 2 is that of Gibbs and Boltzmann: \emph{deriving} all the details of thermodynamics by (a) recognizing that a thermal system is governed by mechanics with very many degrees of freedom, (b) making some simple statistical assumptions about this mechanical system, and (c) identifying certain statistical/mechanical quantities with e.g. temperature and entropy. Level 3 is then to \emph{justify} the statistical assumptions of Level 2, i.e. to explain why thermal equilibrium is in fact such a ubiquitous phenomenon. This has remained a difficult problem for over a century, with only partial progress so far. 

In the QM story, the analog of mechanics is the (deterministic) picture of Hilbert spaces, wavefunctions and unitary evolution. The analog of thermodynamics is the (probabilistic) postulates describing the outcomes of measurements. The latter can be summarized as:
\begin{enumerate}[label=\textbf{\Alph*.}]
 \item An observable property of a quantum system is represented by a Hermitian operator.
 \item A measurement will always yield one of this operator's eigenvalues. 
 \item For a system prepared in a state $\ket{\psi}$, an eigenvalue $\lambda$ will be observed with probability $\braket{\psi|\hat P_\lambda|\psi}$, where $\hat P_\lambda$ is the projector onto the corresponding eigenspace. 
 \item Afterwards, the system appears to ``collapse'' into the observed eigenspace, such that the state $\ket{\psi}$ is replaced by a normalized multiple of $\hat P_\lambda\!\ket{\psi}$.
\end{enumerate}
The fathers of QM, like those of thermodynamics, were at Level 1: they discovered these rules, but couldn't derive them from more fundamental principles. Most modern textbooks stick to Level 1 as well. Reaching the higher levels has been the job of the ``quantum foundations'' community. Unfortunately, this community's bar for consensus is somewhere at Level 3: Boltzmann would never have passed it. As a (somewhat extreme) case in point, see \cite{Kastner}, which literally criticizes some work for being as flawed as Boltzmann's understanding of the Second Law. 
 
The claim of this paper is that the understanding of quantum measurement is now at Level 2: that the above four postulates can be derived by applying wavefunctions and unitary evolution to the relevant macroscopic systems, with some simple assumptions about these systems' behavior, and some rules for relating the formalism to reality (of course, for the exercise to be meaningful, these need to be more basic than Postulates \textbf{A}-\textbf{D} themselves). As with statistical mechanics, our assumptions will appear quite plausible and innocent, but lead to surprisingly strong conclusions when taken seriously. On the other hand, we make no attempt to justify them rigorously: Level 3 remains distant. 

All of the building blocks in our construction are in the published literature. Some date to the 1980's \cite{Zurek:1982ii,Joos:1984uk,Farhi:1989pm}, others to the 2000's \cite{Zurek:2003zz,Zurek:2007bs}. For treatments of some aspects on a more sophisticated level, see e.g. \cite{Deutsch:1999gs,Wallace:ProveBorn,Zurek:2012ny}. For an especially delightful variation on the derivation of Postulate \textbf{C}, see \cite{AharonovReznik}.

\section{Our assumptions} \label{sec:assumptions}

We model a \emph{quantum experiment} as a closed quantum system, with a Hilbert space $\calH$ and a \emph{microstate} $\ket{\Psi}\in\calH$, which undergoes unitary time evolution. This system is \emph{composite}, i.e. its Hilbert space decomposes as $\calH = \bigotimes_i\calH^{(i)}$, where $\calH^{(i)}$ are the Hilbert spaces of subsystems. The latter come in three kinds: \emph{microsystems} $s$, with low-dimensional Hilbert spaces $\calH^{(s)}$, whose quantum states we wish to measure; \emph{macrosystems} $S$, such as \emph{measuring apparatuses} and \emph{grad students}; and finally, the \emph{environment} $\calE$. The Hilbert spaces $\calH^{(S)},\calH^{(\calE)}$ of macrosystems $S$ and the environment $\calE$ are very high-dimensional. As a result, almost any two vectors inside them are almost completely orthogonal.

The detailed microstates of the environment and macrosystems are, in practice, inaccessible. Instead, what we observe are certain \emph{macrostates} $\gamma$ of the macrosystems $S$. These need to satisfy some physical requirements. First, they must be stable under the unavoidable interactions between $S$ and the environment $\calE$. Second, we should be able to view a collection of macrosystems $(S_1,\dots,S_k)$ as one composite macrosystem, with joint macrostates $(\gamma_1,\dots,\gamma_k)$. Somehow, these properties need to be modeled within the Hilbert-space framework. We adopt the following simple model.
\begin{axiom}
  A macrostate $\gamma$ of a macrosystem $S$ is defined by a subspace $\calH_\gamma^{(S)}\subset\calH^{(S)}$, whose dimension is still very large, but much smaller than that of $\calH^{(S)}$. A joint macrostate $(\gamma_1,\dots,\gamma_k)$ of a composite macrosystem $(S_1,\dots,S_k)$ is defined by the direct product $\bigotimes_{i=1}^k\calH^{(S_i)}_{\gamma_i}$. The stability of macrostates under interactions with the environment is encoded in the time-evolution rule:
  \begin{align}
   \calH^{(S)}_\gamma\otimes\calH^{(\calE)} \longmapsto \ \calH^{(S)}_\gamma\otimes\calH^{(\calE)} \ . \label{eq:S_E}
  \end{align}
  The notation here indicates that states in the LHS Hilbert space evolve into states in the RHS Hilbert space (in this case, the same space).
  We say that macrosystem $S$ realizes macrostate $\gamma$, if the total microstate $\ket{\Psi}\in\calH^{(S)}\otimes \calH_{\text{rest}}$ (where $\calH_{\text{rest}}$ denotes the Hilbert space of all other subsystems) satisfies $\ket{\Psi}\in\calH^{(S)}_\gamma\otimes \calH_{\text{rest}}$. Similarly, if $\ket{\Psi}$ will satisfy this condition in the \emph{future}, we say that the $\gamma$ is \emph{predicted}. 
\end{axiom}
Let us now make a few comments. First, the subspaces $\calH_\gamma^{(S)}$ for different macrostates $\gamma$ are almost completely orthogonal to each other. This isn't an assumption, but a generic property of relatively small subspaces in a very high-dimensional space $\calH^{(S)}$. Second, while an interaction of the form \eqref{eq:S_E} preserves the product space $\calH^{(S)}_\gamma\otimes\calH^{(\calE)}$, it can of course change (and will typically increase) the entaglement between $S$ and $\calE$: in particular, product states $\ket{S}\otimes\ket{\calE}$ will not be preserved. Third, note that we only defined here \emph{deterministic} predictions. Of course, these aren't always possible, since the microstate won't always satisfy $\ket{\Psi}\in\calH^{(S)}_\gamma\otimes \calH_{\text{rest}}$ for some $\gamma$. In particular, we can have a situation in which the macrostate of \emph{some} macrosystems $S$ can be predicted, but not that of others. Finally, since this is physics, we must acknowledge that perfect deterministic predictions are at best an idealization. It's therefore good to have a notion of a prediction that \emph{becomes} deterministic in an idealized \emph{limit}:
\begin{axiom_refinement}
  Consider a series of experiments, labeled by positive integers $N$. If the total microstate in each experiment takes the form $\ket{\Psi} = \ket{\Phi} + \ket{\chi}$, where $\ket{\Phi}\in\calH^{(S)}_\gamma \otimes\calH_{\text{rest}}$ and $\lim_{N\rightarrow\infty}\braket{\chi|\chi} = 0$, we say that the macrostate $\gamma$ for macrosystem $S$ is realized (or predicted) in the limit $N\rightarrow\infty$.
\end{axiom_refinement}
Our final set of assumptions specifies the behavior of the particular macrosystems in the experiment. A \emph{measuring apparatus} $A$ is a special kind of macrosystem, capable of measuring a microsystem $s$. Specifically, $A$ has a ``ready to measure'' macrostate, denoted by $\emptyset$, as well as some macrostates $\lambda$ which represent various measurement outcomes. The change in $A$'s macrostate upon measuring $\lambda$ is realized by a unitary (i.e. inner-product-preserving) evolution operator $\hat U^{(A)}_\lambda\!:\,\calH^{(A)}_\emptyset\!\rightarrow \calH^{(A)}_\lambda$. This evolution is triggered when $A$ encounters the microsystem $s$ in certain special states $\ket{s_\lambda}\in\calH^{(s)}$ (there may be more than one such state for every $\lambda$). All in all, we assume a time evolution of the form:
\begin{gather}
 \ket{s_\lambda}\otimes\ket{A_\emptyset} \longmapsto \ket{s_\lambda}\otimes\left(\hat U^{(A)}_\lambda\!\ket{A_\emptyset}\right) , \label{eq:s_A} \\
 \text{for all}\ \ket{A_\emptyset}\in\calH^{(A)}_\emptyset \ . \nonumber
\end{gather}
A \emph{grad student} $G$ is yet another special macrosystem, capable of drawing conclusions from the measurement outcomes of apparatuses. Like an apparatus, $G$ has a ``ready to work'' macrostate, denoted by $\emptyset$, as well as some ``outcome'' macrostates (which may be physically realized e.g. by markings in a lab notebook). $G$ changes her macrostate upon encountering a set of apparatuses $(A_1,\dots,A_k)$, according to some function $f$ of their measurement outcomes $(\lambda_1,\dots,\lambda_k)$. This behavior is captured by time evolution of the form:
\begin{align}
 \left(\bigotimes_{i=1}^k\calH^{(A_i)}_{\lambda_i}\right) \otimes \calH^{(G)}_\emptyset \longmapsto \left(\bigotimes_{i=1}^k\calH^{(A_i)}_{\lambda_i}\right) \otimes \calH^{(G)}_{f(\lambda_1,\dots,\lambda_k)} \ . \label{eq:A_G}
\end{align}
For our purposes, it will be sufficient to consider boolean-values functions $f$. 

\section{Deriving the measurement postulates} \label{sec:derivation}

We'll now derive the measurement postulates, as listed in section \ref{sec:intro}, from the assumptions of section \ref{sec:assumptions}. In our derivations, the environment $\calE$ with its interactions \eqref{eq:S_E} won't play an active role: its only function is to keep us reasonably realistic, and to motivate the structure of the Macrostate Axiom above. In the following, we'll omit $\calE$ and its states for brevity; the reader can verify that restoring them won't affect the conclusions.

\subsection{An observable is a Hermitian operator}

A Hermitian operator is defined by two elements: a collection of mutually orthogonal eigenspaces, and an assignment of real eigenvalues to these eigenspaces. The latter is purely cosmetic: measurement outcomes can just as well be labeled by complex numbers, or by some non-numeric tokens. Thus, the true content of Postulate \textbf{A} is: \emph{states $\ket{s_\lambda}$ of the microsystem $s$ that lead to measurement outcome $\lambda$ form a subspace of $\calH^{(s)}$, and these subspaces are mutually orthogonal}. In other words, there exist projectors $\hat P_\lambda$ onto the relevant subspaces of $\calH^{(s)}$, which satisfy:
\begin{align}
 \hat P_\lambda\hat P_{\lambda'} = \delta_{\lambda\lambda'}\hat P_\lambda \ . \label{eq:ortho_subspaces}
\end{align}
We can now verify that this indeed follows from the assumptions of section \ref{sec:assumptions}. First, consider two states $\ket{s_\lambda},\ket{\tilde s_\lambda}$ that both lead to the same measurement outcome $\lambda$, in the sense of eq. \eqref{eq:s_A}. From the linearity of time evolution, it follows that any superposition $\alpha\ket{s_\lambda}+\beta\ket{\tilde s_\lambda}$ will also satisfy eq. \eqref{eq:s_A}. Thus, the states that lead to measurement outcome $\lambda$ indeed form a subspace. Next, consider two states $\ket{s_\lambda},\ket{s_{\lambda'}}$ that lead to different measurement outcomes $\lambda\neq\lambda'$. Let us demonstrate that these are orthogonal. We restrict attention to a particular initial state $\ket{A_\emptyset}\in\calH^{(A)}_\emptyset$ of the apparatus, and denote:
\begin{align}
 \hat U^{(A)}_\lambda\!\ket{A_\emptyset}\equiv\ket{A_\lambda} \ ; \quad \hat U^{(A)}_{\lambda'}\!\ket{A_\emptyset}\equiv\ket{A_{\lambda'}} \ .
\end{align}
Then the measurement process \eqref{eq:s_A} for $s$ prepared in the state $\ket{s_\lambda}$ or $\ket{s_{\lambda'}}$ reads:
\begin{align}
 \begin{split}
  \ket{s_\lambda}\otimes\ket{A_\emptyset} &\longmapsto \ket{s_\lambda}\otimes\ket{A_\lambda} \ ; \\ 
  \ket{s_{\lambda'}}\otimes\ket{A_\emptyset} &\longmapsto \ket{s_{\lambda'}}\otimes\ket{A_{\lambda'}} .
 \end{split} \label{eq:lambda_lambda'}
\end{align}
Since time evolution is unitary, the inner product of the initial states in \eqref{eq:lambda_lambda'} should equal that of the final states:
\begin{align}
 \braket{s_\lambda|s_{\lambda'}}\braket{A_\emptyset|A_\emptyset} = \braket{s_\lambda|s_{\lambda'}}\braket{A_\lambda|A_{\lambda'}} \ . \label{eq:inner_products}
\end{align}
Now, on the LHS, we have $\braket{A_\emptyset|A_\emptyset}=1$, while on the RHS, $\braket{A_\lambda|A_{\lambda'}}$ is almost certainly close to zero, since $\calH^{(A)}_\lambda$ and $\calH^{(A)}_{\lambda'}$ are almost completely orthogonal. Therefore, \eqref{eq:inner_products} can only hold if $\braket{s_\lambda|s_{\lambda'}}=0$, as we wanted to show. 

Note that for this derivation, the orthogonality of $\calH^{(A)}_\lambda$ and $\calH^{(A)}_{\lambda'}$ (due to the macroscopic nature of the apparatus) isn't really needed. It's enough to have $\left|\braket{A_\lambda|A_{\lambda'}}\right|<1$, which merely requires that the states $\ket{A_\lambda},\ket{A_{\lambda'}}$ \emph{do not coincide}. Thus, Postulate \textbf{A} is a robust consequence of any interaction of the form \eqref{eq:lambda_lambda'}, in which the ``measuring apparatus'' can be just another microscopic system \cite{Zurek:2007bs}. The orthogonality of $\calH^{(A)}_\lambda,\calH^{(A)}_{\lambda'}$ will become important later, in our derivation of Postulate \textbf{D}.

\subsection{We always observe one of the eigenvalues}

Consider a microsystem $s$ prepared in a superposition $\ket{\psi}\in\ket{1}\oplus\ket{2}$ of two states $\ket{1},\ket{2}$ that lead to measurement outcomes 1,2 by an apparatus $A$. Postulate \textbf{B} states that ``the measurement will always yield one of the outcomes 1,2''. To evaluate this claim, we will employ a grad student $G$. Her job is to observe the apparatus, and mark in her notebook which of the following two statements is true:
\begin{itemize}
	\item Postulate \textbf{B} is true: the apparatus' needle is pointing at one of the allowed values 1,2.
	\item Postulate \textbf{B} is false: the apparatus' needle is pointing at some intermediate value, or there are suddenly two needles pointing at both 1 and 2, or instead of a needle I see a cloud of probability, or I've been thrust out of my ordinary experience into a psychedelic vision of quantum reality, under the guidance of a talking cactus. 
\end{itemize}
Let's label the corresponding macrostates of $G$ as ``B-true'' and ``B-false''. Now, for $s$ prepared in one of the two eigenstates $\ket{1},\ket{2}$, the time evolution will take the form:
\begin{align}
 \begin{split}
   \ket{1}\otimes\calH^{(A)}_\emptyset\otimes\calH^{(G)}_\emptyset &\longmapsto \ket{1}\otimes\calH^{(A)}_1\otimes\calH^{(G)}_{\text{B-true}} \ ; \\ 
   \ket{2}\otimes\calH^{(A)}_\emptyset\otimes\calH^{(G)}_\emptyset &\longmapsto \ket{2}\otimes\calH^{(A)}_2\otimes\calH^{(G)}_{\text{B-true}} \ .
 \end{split}
\end{align}
Since time evolution is linear, for $s$ prepared in a superposition, we get:
\begin{align}
 \begin{split}
   &\big(\!\ket{1} \oplus \ket{2}\!\big)\otimes\calH^{(A)}_\emptyset\otimes\calH^{(G)}_\emptyset \\
   &\ \longmapsto \left(\left(\ket{1}\otimes\calH^{(A)}_1\right) \oplus \left(\ket{2}\otimes\calH^{(A)}_2\right) \right)\otimes\calH^{(G)}_{\text{B-true}} \ .
 \end{split}
\end{align}
Thus, we predict deterministically that $G$ will attest to seeing one of the outcomes 1,2! The extension to more than two outcomes is trivial. Obviously, $G$'s notebook in this argument can be replaced by \emph{any} macrosystem that can evaluate the truth or falsehood of Postulate \textbf{B}, such as e.g. the mind of a postdoc at the same lab.

\subsection{The Born rule for probabilities} \label{sec:derivation:Born}

Consider again a microsystem $s$ prepared in a state $\ket{\psi}$, to be measured by an apparatus $A$. For simplicity, assume only two possible measurement outcomes 1,2. We can then decompose $\ket{\psi}$ into the corresponding eigenspaces, as:
\begin{align}
 \ket{\psi} = \hat P_1\!\ket{\psi} + \hat P_2\!\ket{\psi} \ . \label{eq:psi_decompose}
\end{align}
The norm-squared of these two terms is respectively $p$ and $1-p$, where we denote:
\begin{align}
 p \equiv \braket{\psi|\hat P_1|\psi} \ . \label{eq:p}
\end{align}
The statement of Postulate \textbf{C} is that \emph{$p$ is the probability} for observing outcome 1. This simply means that, if we repeat the experiment $N$ times, the relative frequency of outcome 1 will tend to the predicted probability $p$ in the limit $N\rightarrow\infty$. This is a prediction that \emph{becomes deterministic in a limit}, as anticipated in the ``refined Macrostate Axiom'' of section \ref{sec:assumptions}. 

Let's make the construction explicit. Consider an experiment with $N$ copies of $s$, all prepared in the state $\ket{\psi}$. These are measured by $N$ copies of the apparatus $A$. To avoid the complications of a continuum limit, we fix some finite error margin $\varepsilon>0$ for the outcomes' relative frequencies. We then assign a grad student $G$ to evaluate the measurement outcomes $(\lambda_1,\dots,\lambda_N)$ of the $N$ apparatuses, as in eq. \eqref{eq:A_G}. The grad student's function $f$ decides whether Postulate \textbf{C} is satisfied, according to:
\begin{align}
 f(\lambda_1,\dots,\lambda_N) = \left\{\begin{array}{lc}
   \displaystyle \text{``C-true''} & \left| \frac{m}{N} - p \right| < \varepsilon \\  
   \text{``C-false''} & \displaystyle \text{else}
   \end{array} \right. \ , \label{eq:Born_student}
\end{align}
where $m$ is the number of ``1'' outcomes among $(\lambda_1,\dots,\lambda_N)$. Our initial state takes the form:
\begin{align}
 \ket{\Psi} \in \left(\left(\hat P_1\!\ket{\psi} + \hat P_2\!\ket{\psi}\right)\otimes\calH^{(A)}_\emptyset\right)^{\otimes N}\otimes\calH^{(G)}_\emptyset \ .
\end{align}
This is a sum of $2^N$ terms, all orthogonal to each other. The norm-squared of each term is $p^m(1-p)^{N-m}$, where $m$ represents the number of $\hat P_1\!\ket{\psi}$ factors in the product, and $N-m$ is the number of $\hat P_2\!\ket{\psi}$ factors. The number of terms with a given value of $m$ is $\binom{N}{m}$. Since time evolution is unitary, the $2^N$ terms will retain their norms and orthogonality throughout the experiment. After measurement \eqref{eq:s_A} by the apparatus, the state evolves into:
\begin{align*}
 \ket{\Psi'} \in \left(\left(\hat P_1\!\ket{\psi}\otimes\calH^{(A)}_1\right) \oplus \left(\hat P_2\!\ket{\psi}\otimes\calH^{(A)}_2\right)\right)^{\otimes N}\otimes\calH^{(G)}_\emptyset \ .
\end{align*}
After evaluation \eqref{eq:A_G} by the grad student, the state evolves into $\ket{\Psi''} = \ket{\Phi} + \ket{\chi}$, where we grouped the $2^N$ terms according to the grad student's outcome:
\begin{align}
 \begin{split}
   \ket{\Phi} &\in \big(\calH^{(s)}\otimes\calH^{(A)}\big)^{\otimes N}\otimes\calH^{(G)}_{\text{C-true}} \ ; \\
   \ket{\chi} &\in \big(\calH^{(s)}\otimes\calH^{(A)}\big)^{\otimes N}\otimes\calH^{(G)}_{\text{C-false}} \ .
 \end{split}
\end{align}
Let's now evaluate the norm of $\ket{\chi}$. Tallying the norm-squared of the orthogonal terms that contribute to it, and using the Gaussian bound on binomial sums, we get:
\begin{align}
 \braket{\chi|\chi} &= \left(\sum_{m \leq N(p - \varepsilon)} + \sum_{m\geq N(p+\varepsilon)}\right)\!\binom{N}{m}p^m(1-p)^{N-m} \nonumber \\
  &\leq 2e^{-2N\varepsilon^2} \underset{N\rightarrow\infty}{\longrightarrow} 0 \ .
\end{align}
Thus, in the $N\rightarrow\infty$ limit, $\ket{\chi}$ vanishes in the sense of the Hilbert-space norm. We can therefore predict deterministically that the grad student's notebook will read ``C-true'', confirming the Born rule. Again, the generalization to more than two outcomes is trivial.

\subsection{Apparent collapse}

Consider again the microsystem $s$ in a state $\ket{\psi}$, measured by $A$ with two possible outcomes $1,2$, with corresponding projectors $\hat P_1,\hat P_2$ on $\calH^{(s)}$. What does it mean to say that ``upon measuring outcome 1, the state of $s$ collapses into a normalized multiple of $\hat P_1\!\ket{\psi}$''? It can only mean that \emph{the outcome of all subsequent measurements will be as if such a collapse occurred}. Consider, then, a second measurement by apparatus $B$, with two possible outcomes $\alpha,\beta$ associated with projectors $\hat P_\alpha,\hat P_\beta$ (the interesting case is of course when $\hat P_\alpha,\hat P_\beta$ don't commute with $\hat P_1,\hat P_2$). The statement of Postulate \textbf{D} is then that \emph{the probability of the sequence of outcomes $(1,\alpha)$ is equal to the probability of outcome 1 for $s$ prepared in the state $\ket{\psi}$, times the probability of outcome $\alpha$ for $s$ prepared in the state $\frac{\hat P_1\!\ket{\psi}}{\sqrt{\braket{\psi|\hat P_1|\psi}}}$, i.e. in a normalized multiple of $\hat P_1\!\ket{\psi}$}. By Postulate \textbf{C}, these latter probabilities are $\braket{\psi|\hat P_1|\psi}$ and $\frac{\braket{\psi|\hat P_1\hat P_\alpha\hat P_1|\psi}}{\braket{\psi|\hat P_1|\psi}}$, respectively. Thus, the statement of Postulate \textbf{D} simplifies into: \emph{the probability of the sequence of outcomes $(1,\alpha)$ is $\braket{\psi|\hat P_1\hat P_\alpha\hat P_1|\psi}$}. We can now easily verify that this is in fact true, by extending the method of section \ref{sec:derivation:Born}.

The process of measuring $s$, first by $A$ and then by $B$, leads to a sum of 4 terms, labeled by the measurement outcomes $\lambda=1,2$ and $\xi=\alpha,\beta$:
\begin{align}
 \begin{split}
   &\ket{\psi}\otimes\calH^{(A)}_\emptyset\otimes\calH^{(B)}_\emptyset \\
   &\quad \longmapsto \bigoplus_{\lambda=1,2} \left(\hat P_\lambda\!\ket{\psi}\otimes\calH^{(A)}_\lambda\right) \otimes\calH^{(B)}_\emptyset \\
   &\quad \longmapsto \bigoplus_{\xi=\alpha,\beta}\bigoplus_{\lambda=1,2} \left(\hat P_\xi \hat P_\lambda\!\ket{\psi}\otimes\calH^{(A)}_\lambda\otimes\calH^{(B)}_\xi \right) \ .
 \end{split} \label{eq:AB_measurement}
\end{align}
The norm-squared of each term is $\braket{\psi|\hat P_\lambda\hat P_\xi\hat P_\lambda|\psi}$. The vectors $\hat P_\xi \hat P_\lambda\!\ket{\psi}$ are orthogonal to each other for different $\xi$, but not for different $\lambda$; however, due to the orthogonality of the  subspaces $\calH^{(A)}_{\lambda}$, the 4 terms in \eqref{eq:AB_measurement} are orthogonal overall. Now, consider $N$ copies of the $(s,A,B)$ setup, each undergoing the process \eqref{eq:AB_measurement}, leading to $4^N$ orthogonal terms overall. We can now add in a graduate student, tasked as in \eqref{eq:Born_student} with evaluating whether the relative frequency of the outcome sequence $(1,\alpha)$ agrees with the corresponding norm-squared $\braket{\psi|\hat P_1\hat P_\alpha\hat P_1|\psi}$, up to some accuracy $\varepsilon$. From here, the argument is identical to that in section \ref{sec:derivation:Born}. In the $N\rightarrow\infty$ limit, the student will deterministically conclude that the relative frequency agrees with the probability that one would expect from wavefunction collapse.

\section*{Acknowledgements}

We are grateful to Philipp Hoehn, Ezra Kassa, David O'Connell and Lev Vaidman for discussions. YN's recent interest in the subject was sparked by the transcript \cite{Coleman:2020put} of a lecture by Sidney Coleman, which was uploaded by Martin Greiter, and advertised by Peter Woit. This work was supported by the Quantum Gravity Unit of the Okinawa Institute of Science and Technology Graduate University (OIST).


\begin{thebibliography}{99}

\bibitem{Kastner} 
Kastner, R.~E.\ 2014, 
``'Einselection' of Pointer Observables: The New H-Theorem?,''
Studies in the History and Philosophy of Modern Physics, 48, 56. 
doi:10.1016/j.shpsb.2014.06.004

\bibitem{Zurek:1982ii}
W.~H.~Zurek,
``Environment induced superselection rules,''
Phys. Rev. D \textbf{26}, 1862-1880 (1982)
doi:10.1103/PhysRevD.26.1862

\bibitem{Joos:1984uk}
E.~Joos and H.~D.~Zeh,
``The Emergence of classical properties through interaction with the environment,''
Z. Phys. B \textbf{59}, 223-243 (1985)
doi:10.1007/BF01725541

\bibitem{Farhi:1989pm}
E.~Farhi, J.~Goldstone and S.~Gutmann,
``How Probability Arises in Quantum Mechanics,''
Annals Phys. \textbf{192}, 368 (1989)
doi:10.1016/0003-4916(89)90141-3

\bibitem{Zurek:2003zz}
W.~H.~Zurek,
``Decoherence, einselection, and the quantum origins of the classical,''
Rev. Mod. Phys. \textbf{75}, 715-775 (2003)
doi:10.1103/RevModPhys.75.715
[arXiv:quant-ph/0105127 [quant-ph]].

\bibitem{Zurek:2007bs}
W.~H.~Zurek,
``Quantum origin of quantum jumps: Breaking of unitary symmetry induced by information transfer and the transition from quantum to classical,''
Phys. Rev. A \textbf{76}, 052110 (2007)
doi:10.1103/PhysRevA.76.052110
[arXiv:quant-ph/0703160 [quant-ph]].

\bibitem{Deutsch:1999gs}
D.~Deutsch,
``Quantum theory of probability and decisions,''
Proc. Roy. Soc. Lond. A \textbf{455}, 3129 (1999)
doi:10.1098/rspa.1999.0443
[arXiv:quant-ph/9906015 [quant-ph]].

\bibitem{Wallace:ProveBorn}
D.~Wallace,
``A formal proof of the Born rule from decision-theoretic assumptions,''
Chapter 8 of ``Many Worlds?: Everett, Quantum Theory, and Reality,'' Oxford University Press (2010)
doi:10.1093/acprof:oso/9780199560561.003.0010
[arXiv:0906.2718 [quant-ph]].

\bibitem{Zurek:2012ny}
W.~H.~Zurek,
``Wave-packet collapse and the core quantum postulates: Discreteness of quantum jumps from unitarity, repeatability, and actionable information,''
Phys. Rev. A \textbf{87}, no.5, 052111 (2013)
doi:10.1103/PhysRevA.87.052111
[arXiv:1212.3245 [quant-ph]].

\bibitem{AharonovReznik}
Y.~Aharonov and B.~Reznik,
``How macroscopic properties dictate microscopic probabilities,''
Phys. Rev. A \textbf{65}, no.5, 052116 (2002)
doi:10.1103/PhysRevA.65.052116
[arXiv:quant-ph/0110093 [quant-ph]].

\bibitem{Coleman:2020put}
S.~Coleman,
``Sidney Coleman's Dirac Lecture ''Quantum Mechanics in Your Face'',''
[arXiv:2011.12671 [physics.hist-ph]].

\end{thebibliography}
\end{document}